\documentclass{jfm}   

\def\bsa#1\esa{\begin{subequations}
\begin{align}#1\end{align} \end{subequations}}

\usepackage{graphicx}
\usepackage{newtxtext}
\usepackage{newtxmath}
\usepackage{natbib}
\usepackage{amsmath}
\usepackage{hyperref}
\usepackage{caption}
\usepackage{subcaption}
\hypersetup{
    colorlinks = true,
    urlcolor   = blue,
    citecolor  = black,
}

\newcommand{\RomanNumeralCaps}[1]
\nolinenumbers

\shortauthor{T. Iida and A. Jensen}
\shorttitle{Homogenized scattering model of water wave attenuation in marginal ice zone}
\title{Homogenized scattering model of water wave attenuation in marginal ice zone}

\author{ Takahito Iida\aff{1}\corresp{email: {iida@naoe.eng.osaka-u.ac.jp} }
\and Atle Jensen\aff{2}
 }

\affiliation{
\aff{1}
Department of Naval Architecture and Ocean Engineering, Osaka University, Osaka 5650871, Japan
\aff{2}
Department of Mathematics, University of Oslo, Oslo 0316, Norway
}

\begin{document}
\maketitle

\begin{abstract}
A theoretical model to explain the scattering process of wave attenuation in a marginal ice zone is developed.
Many field observations offer wave energy decay in the form of exponential function with distance, and this is justified through the complex wave number for the dissipation process.
On the other hand, such a mechanism is not explicitly proven for the scattering process.
To explain this, we consider a periodic array of ice floes, where the floe is modeled by a vertical rigid cylinder.
Using a homogenization technique, a homogenized free surface equivalent to the array is obtained.
Then, we show that a dispersion relation of the homogenized free surface waves makes all wave numbers complex.
As a result, the exponential energy decay in the scattering process is demonstrated.
Although our model is obtained using many simplifications, it reproduces consistent tendencies with both existing field observations and numerical simulations; the wave attenuation coefficient for the deep sea is proportional to the ice concentration and the wave number for open water waves, and the coefficient is bigger as the radius and draft of the floe become larger or the wave period is smaller.

\end{abstract}
\begin{keywords}
Water waves, marginal ice zone, wave attenuation, scattering process, dispersion relation, homogenization
\end{keywords}

\section{Introduction}
\label{intr}

Accurate global wave hindcast is essential to utilize the ocean space.
Such a wave hindcast is based on the energy transport equation \cite[e.g. WAVEWATCH III{\textregistered};][]{wavewatch2019}, and a source term is represented by a linear sum of some components, such as wind-wave interaction, nonlinear wave-wave interaction, wave breaking (white-capping), and wave-ice interaction.
As wave-ice interaction is likely to contribute one of the important roles to global dynamics \cite[]{stroeve2007, squire2020}, an improvement of numerical and theoretical models of wave-ice interaction is demanded \cite[e.g.][]{thomson2018}.

Wave-ice interaction is especially of great importance in a marginal ice zone where numerous pieces of compact ice are floating on the surface of water.
Since wave energy exponentially decays with distance \cite[]{robin1963, wadhams1988}, wave-ice interaction is described by a wave attenuation coefficient.
This attenuation is known as a result of two processes, i.e. energy dissipation and scattering \cite[see the latest review by][]{squire2020}.
The dissipation covers various factors, such as viscosity, inelastic collisions, overwash, wave breaking, ice breaking, and so on, and (mechanical) energy is not conserved in this process.
Two-layer viscous models are often used to explain the dissipation process \cite[e.g.][]{de2002,wang2010,sutherland2019}.
The dispersion relation of these models offers a complex wave number \cite[]{keller1998,de2002}.
This indicates that an imaginary part of the wave number yields exponential decay of wave amplitude and energy with distance.
Such a dispersion relation is incorporated in global wave hindcasts to determine the wave attenuation coefficient \cite[e.g.][]{wavewatch2019}.
Despite their efforts, further improvement of models is still required from the view of overcoming homogeneous linear assumptions \cite[]{squire2020}.
Therefore, more investigations into nonlinear dynamics and modeling them are necessary.
For example, it is reported that the collision of ice floes induces turbulence, and it results in energy dissipation \cite[]{loken2022}.

The energy scattering process, on the other hand, conserves energy; propagating wave energy is just redistributed by a boundary of ice, i.e. the propagating energy decays due to wave reflection (or scattering) by ice.
\cite{perrie1996} incorporated a motion of a rigid ice floe into the energy transport equation to calculate a wave attenuation coefficient.
Except for their work, an elastic model is preferred, developed, and widely used \cite[e.g.][]{squire1995, kohout2008,bennetts2009}.
Recently, sophisticated simulation methods of a very large number of ice floes are also established \cite[][]{bennetts2010,montiel2016}.
Most scattering occurs when ice length is comparable to water wavelength \cite[][]{squire2020}.
In addition, the wave attenuation coefficient increases as the wave period becomes shorter \cite[e.g.][]{li2017}. 
On the other hand, the elastic response of the floe appears when the ice length is bigger than the characteristic length \cite[][]{suzuki1996}.
Interestingly, these parameters (i.e. length of the floe, wavelength, and characteristic length) are of a similar order.
In that sense, the elastic mode is not negligible.
However, the scattering process itself is based on reflection at the edge of the ice floe; it occurs regardless of the rigid or elastic modeling.
Although a thin elastic plate has two complex wave numbers, it also has a real wave number.
This means that wave energy does not vanish far from the inlet ice edge if the ice plate is a semi-infinite length.
Therefore, the elastic model itself does not explain the mechanism of the exponential decay of energy.
In addition, such a model needs numerical simulations, and propagating energy is calculated from the transmission coefficients \cite[e.g.][]{kohout2008,bennetts2009}.
They assumed the exponential energy decay based on the field observations, and the wave attenuation coefficient is fitted.
Of course, as far as seeing results, this assumption works well.
However, this is not theoretically proven, unlike the dissipation process.
Besides, it is not easy to incorporate these numerical simulations into wave hindcasts.
For example, WAVEWATCH III{\textregistered} uses fit curves of their simulations as empirical options \cite[][]{wavewatch2019}.

In this paper, we propose a new scattering model to describe a water wave attenuation in the marginal ice zone.
Our aim is not to establish a very accurate simulation model.
Instead, we aim to present a theoretical model to prove exponential energy decay using a simplified formulation.
We believe this facilitates the understanding of wave attenuation due to the scattering process.
We assume an ice floe is a vertical rigid cylinder, and a marginal ice zone is modeled by a periodic array of such a cylinder.
Using a homogenization \cite[][]{garnaud2009}, a homogenized free surface equivalent to the periodic array of ice floe is obtained.
This makes all wave numbers complex, and this results in wave attenuation in the form of an exponential function with distance.
Parameter studies are carried out to understand the relation between the wave attenuation coefficient and physical parameters.
We also show some simulation results compared with existing field observations and simulation models.


\begin{figure}
     \centering
     \begin{subfigure}{0.2\textwidth}
         \centering
         \includegraphics[width=\textwidth]{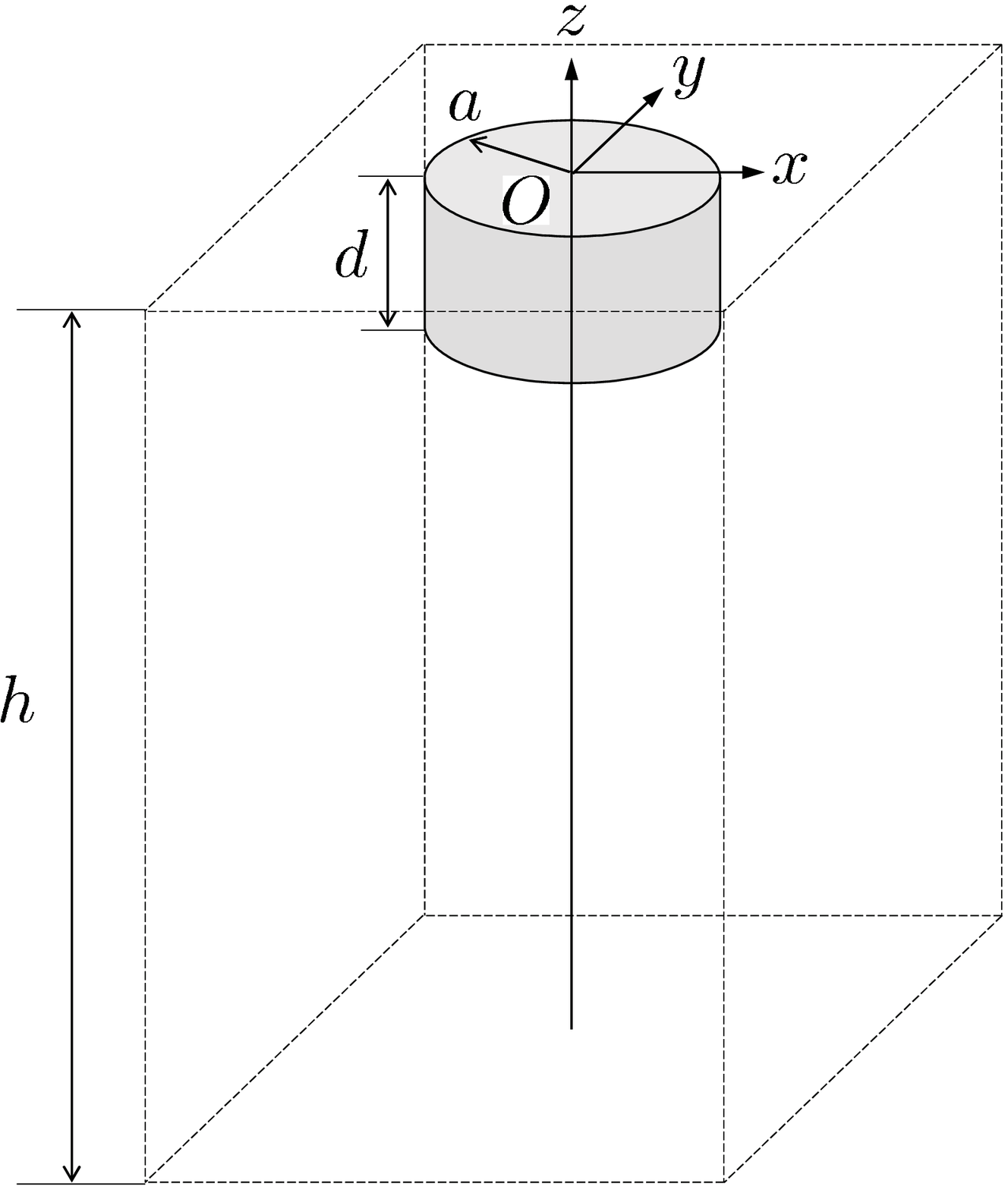}
         \caption{}
        \label{fig1a}
     \end{subfigure}
     \begin{subfigure}{0.39\textwidth}
         \centering
         \includegraphics[width=\textwidth]{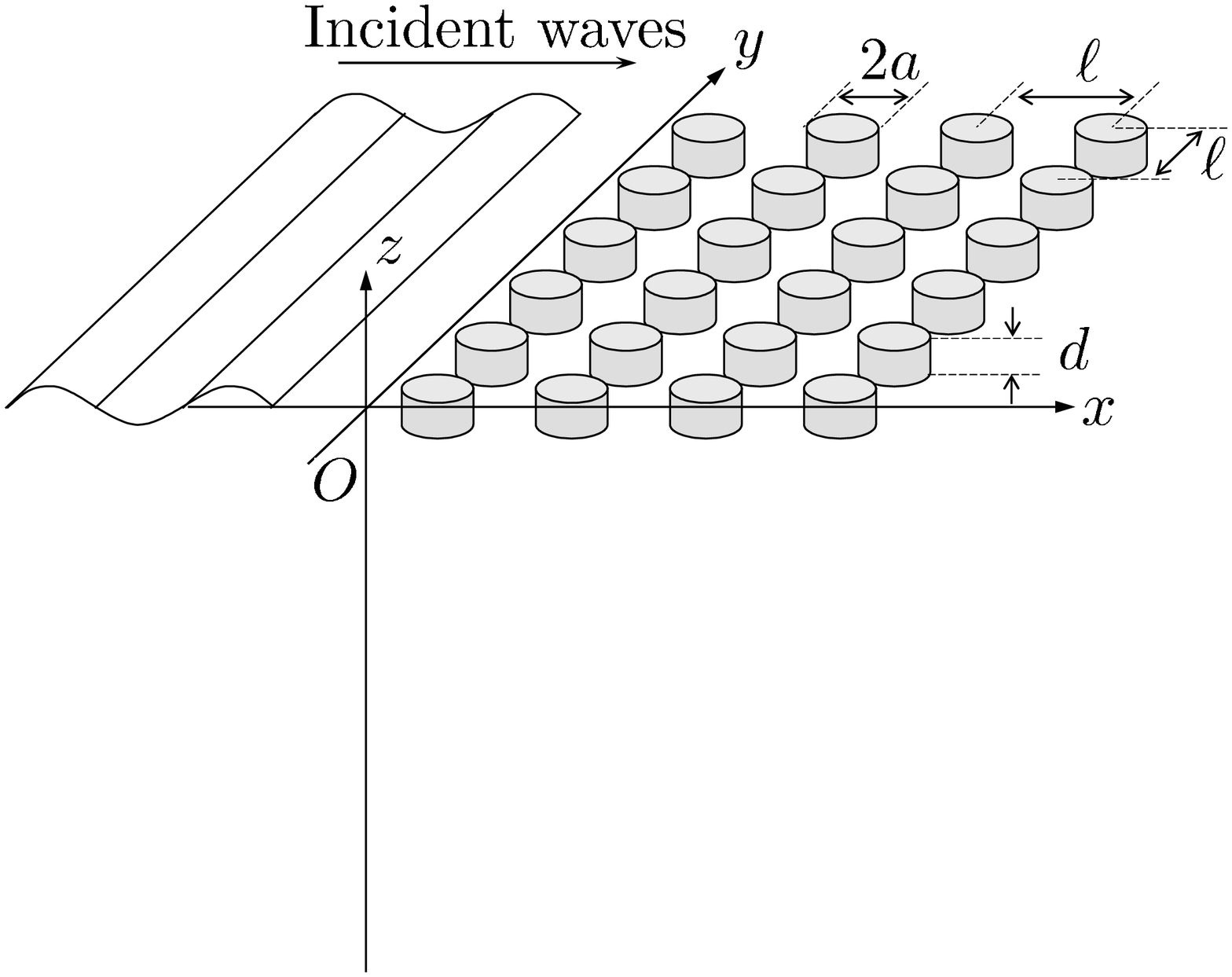}
         \caption{}
        \label{fig1b}
     \end{subfigure}
     \begin{subfigure}{0.39\textwidth}
         \centering
         \includegraphics[width=\textwidth]{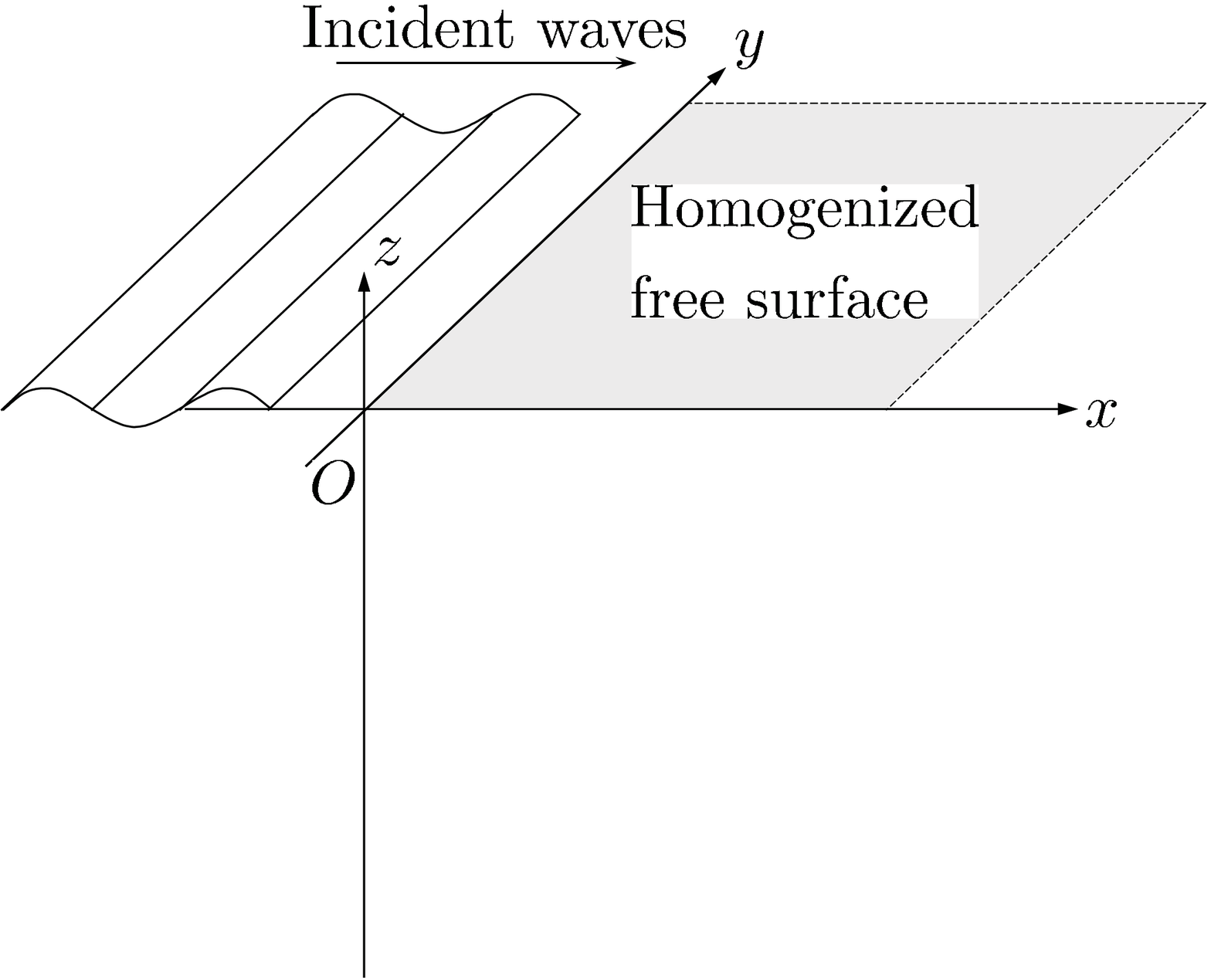}
         \caption{}
        \label{fig1c}
     \end{subfigure}
 \caption{Problem descriptions and concepts. (a) The local coordinate system of a single ice floe. An ice floe is modeled by a floating vertical cylinder with radius $a$ and draft $d$. (b) The global coordinate system of marginal ice zone where marginal ice is modeled by a periodic array of cylinders. Distance between floes is $\ell$. (c) Concept of a homogenized free surface. The periodic array of ice floes in Fig.\ref{fig1b} is replaced with the homogenized free surface. This paper aims to find such a free surface.}
        \label{fig1}
\end{figure}

\section{Theoretical description}
We consider a marginal ice zone consisting of discrete small ice floes with a low concentration ratio.
For such a case, the process of wave attenuation is mainly described by the wave scattering \cite[]{squire1995}.
In the present paper, a new theoretical model is proposed to facilitate the understanding of wave attenuation due to the scattering process.
The problem schematics and concepts are described in Fig. \ref{fig1}.
A small ice floe is modeled by a truncated vertical rigid cylinder floating on the free surface as shown in Fig. \ref{fig1a}.
Furthermore, the marginal ice zone is modeled by a periodic array of such an ice floe as in Fig. \ref{fig1b}.
Here, we replace the array with a homogenized free surface equivalent to the array as in Fig. \ref{fig1c}.
Using this homogenized free surface, characteristics of waves in the marginal ice zone are discussed.
The following subsections describe the details of the proposed theories.


\subsection{Boundary value problem of single ice floe in waves}\label{sec2.1}
Firstly, a boundary value problem of a single ice floe in waves is briefly reviewed.
We consider the three-dimensional coordinate system $O$-$xyz$ where $z=0$ plane denotes the undisturbed free surface of water, and vertically upward is defined by positive $z$ (see Fig. \ref{fig1a}).
The sea bottom is assumed flat at $z=-h$.
As the simple modeling, the shape of the ice floe is assumed a vertical cylinder floating on the free surface, of which radius and draft are $a$ and $d$.
Here, only the vertical motion (heave motion) of the floe is considered.
Besides, the floe is rigid, and an elastic response is not considered.
We formulate the boundary value problem of the floe based on the potential flow theory \cite[e.g.][]{newman2018}; incompressible and inviscid fluid with the irrotational motion of a fluid are assumed.
Furthermore, long-crested plane waves with a circular frequency $\omega$ are considered.
Wave amplitude and resultant floe's motion are sufficiently smaller than wavelength $\lambda$, and thus the linearization is applied.
This results in the time-harmonic solutions, such as velocity potential $\Phi(\mbox{\boldmath $x$},t)={\rm Re}[\phi(\mbox{\boldmath $x$})\exp(i\omega t)]$, wave elevation $A(\mbox{\boldmath $x$},t)={\rm Re}[\zeta(\mbox{\boldmath $x$})\exp(i\omega t)]$, and heave motion of the floe $x_3(t)={\rm Re}[X_3\exp(i\omega t)]$.
Then, the linearized boundary value problem in a frequency domain is given as
\begin{align}
& \nabla^2 \phi=0 & (-h\le z\le 0), \label{bvp:01}\\
&\displaystyle \frac{\partial \phi}{\partial z}=0 & (z=-h), \label{bvp:02}\\
&\displaystyle \frac{\partial \phi}{\partial z}=i\omega X_3 & (z=-d), \label{bvp:03}
\end{align}
\begin{subequations}\vspace{-2\abovedisplayskip}
\begin{align}
&\displaystyle \frac{\partial \phi}{\partial z}=i\omega \zeta &\quad\quad\quad (z=0), \label{bvp:04a}\\
&\displaystyle \zeta=-\frac{i\omega}{g}\phi & (z=0), \label{bvp:04b}\\
&\displaystyle \frac{\partial \phi}{\partial z}=\frac{\omega^2}{g}\phi& (z=0), \label{bvp:04c}
\end{align}
\end{subequations}
where $g$ is the gravitational acceleration.
Here, (\ref{bvp:01}) is the Laplace equation that governs the fluid domain, (\ref{bvp:02}) is the sea bottom condition, and (\ref{bvp:03}) is the floe bottom condition of which fluid velocity coincides with floe's velocity.
Equations (\ref{bvp:04a}) and (\ref{bvp:04b}) are kinematic and dynamic conditions of the free surface, respectively.
Combining (\ref{bvp:04a}) and (\ref{bvp:04b}), the linearized free surface condition (\ref{bvp:04c}) is obtained.
In addition, no flux condition is imposed for a side wall.
This problem is classical, and this can be solved by any numerical simulation methods, such as an eigenfunction matching method \cite[]{miles1968,garrett1971}, boundary element method \cite[]{lee1988}, and CFD.
Here, the eigenfunction matching method is employed to obtain the motion of the single ice floe.
Note that we solve the following equation of motion
\begin{align}
\bigg[-\bigg(1+\frac{A_{33}}{\rho \pi a^2 d}\bigg)+i \frac{B_{33}}{\omega \rho \pi a^2 d}+\frac{g}{\omega^2 d}\bigg]\frac{X_3}{\zeta_0}=\frac{g}{\omega^2 d}\frac{E_3}{\pi a^2 \rho g  \zeta_0},\label{bvp:05}
\end{align}
where $\zeta_0$ is incident wave amplitude, $A_{33}$ is the added mass, $B_{33}$ is the damping coefficient, and $E_3$ is the wave exciting force.

\subsection{Homogenized boundary value problem of waves in marginal ice zone}
We assume that the marginal ice zone is modeled by an array of the floes periodically-arranged in distance $\ell$ (see Fig. \ref{fig1b}).
The floe is represented by the floating vertical cylinder as described in \S\ref{sec2.1}, and all floes' radii and drafts are uniforms.
The radius, draft, and distance are smaller than the wavelength, or at most in the same order as the wavelength.
In this subsection, we aim to model a homogenized free surface consisting of the free surface of the water and the surface of floes as in Fig. \ref{fig1c}.
This idea is inspired by \cite{garnaud2009} and \cite{mei2012}; they developed a model for an array of point absorbers, but they implied the applicability of their model to a problem of small ice floes.
Nevertheless, the point absorber has an energy extraction term which is assumed $O(1)$, and hydrodynamic forces are ignored as $O(\varepsilon)$.
In the case of ice floes, on the other hand, hydrodynamic forces must be taken into account, and this influence does not appear in the leading order solution.
Therefore, we shall modify the theory to bring it in line with our problem.

We focus on one floe in the array.
The floe is surrounded by a unit cell of which the horizontal squire area is $\ell^2$.
To obtain the homogenized free surface, it is assumed that the motion of the ice floe imitates a wave elevation, namely, a pseudo-wave elevation.
Then, a new wave amplitude $\eta$ in the local coordinate is defined as
\begin{align}
\eta(r,\theta)=
\left\{
\begin{array}{ll}
\vspace{1mm}
\displaystyle \zeta &(r>a; \text{on water surface})\\
\displaystyle X_3&(r\le a; \text{on floe surface})
\end{array}\right.,
 \label{pb:01}
\end{align}
where $(r,\theta)$ is a horizontal coordinate from the center of the floe (local coordinate system).
Firstly, boundary conditions of the pseudo-wave elevation are considered.
Assuming the small draft of the floe, the floe bottom condition (\ref{bvp:03}) is approximated by the Taylor-series expansion at the undisturbed free surface $z=0$ as
\begin{align}
&\displaystyle i\omega X_3=\frac{\partial \phi}{\partial z}+O(\phi d)&(z=0). \label{pb:02}
\end{align}
We call (\ref{pb:02}) a pseudo-kinematic condition from the analogy of the kinematic condition (\ref{bvp:04a}).
Furthermore, we assume the floe's motion is represented by the product of non-dimensional motion amplitude and wave amplitude, i.e.
\begin{align}
\displaystyle X_3=X^*_3\zeta,
\label{pb:03}
\end{align}
where non-dimensional motion amplitude holds $X^*_3=X_3/\zeta_0$ which is found in (\ref{bvp:05}).
Note that non-dimensional motion amplitude $X^*_3$ is calculated by (\ref{bvp:05}), and this value is the same for all floes.
On the other hand, the dimensional motion amplitude $X_3$ of each floe depends on its global position.
Applying the dynamic condition for free surface waves (\ref{bvp:04b}) into (\ref{pb:03}), we get
\begin{align}
&\displaystyle X_3=-\frac{i\omega}{g}X^*_3\phi&(z=0),
\label{pb:04}
\end{align}
where (\ref{pb:04}) is called a pseudo-dynamic condition.
Combining (\ref{bvp:04a}) and (\ref{pb:02}), a new kinematic condition is given as
\begin{align}
&\displaystyle \frac{\partial \phi}{\partial z}=i\omega \eta&(z=0). \label{pb:05}
\end{align}
Similarly, (\ref{bvp:04b}) and (\ref{pb:04}) yield a new dynamic condition as
\begin{align}
&\displaystyle \eta=-\frac{i\omega}{g}f(r)\phi & (z=0), \label{pb:06}
\end{align}
where
\begin{align}
f(r)=
\left\{
\begin{array}{ll}
\vspace{1mm}
\displaystyle 1 &(r>a)\\
\displaystyle X_3^*&(r\le a)
\end{array}\right..
 \label{pb:07}
\end{align}

We further deform (\ref{pb:06}) by averaging wave amplitude over the surface of the unit cell.
Mean wave amplitude is calculated by
\begin{align}
\displaystyle \overline{\eta}=\frac{1}{\ell^2}\iint_{\Delta S_F}\eta ds=-\frac{i\omega}{g\ell^2}\iint_{\Delta S_F}f(r)\phi ds, \label{pb:08}
\end{align}
where $\Delta S_F$ denotes the surface boundary of the cell.
Considering periodicity for the unit cell's surrounding boundary, velocity potential is independent of local coordinate  \cite[see][]{garnaud2009}.
It facilitates the calculation of (\ref{pb:08}) as
\begin{align}
\displaystyle \overline{\eta}&\approx -\frac{i\omega}{g\ell^2}\phi\iint_{\Delta S_F}f(r)ds=-\frac{i\omega}{g}\phi \frac{1}{\ell^2}[(\ell^2-S)+X_3^*S]\notag\\
&=-\frac{i\omega}{g}[1+\psi(X_3^*-1)]\phi,\label{pb:09}
\end{align}
where $S=\pi a^2$ is waterplane area of the floe and $\psi=S/\ell^2$ is an ice concentration ratio (a.k.a. filling ratio).
Note that (\ref{pb:09}) is valid not only for a vertical cylinder but also for other geometries (such as a rectangular plate).
When the cylinder is considered, the maximum concentration ratio is $\psi=\pi/4$.

Summarizing the above boundary conditions, the homogenized boundary value problem of wave propagation in the marginal ice zone is formulated as
\begin{align}
& \nabla^2 \phi=0 &\quad\quad\quad\quad\quad (-h\le z\le 0), \label{pb:10}\\
&\displaystyle \frac{\partial \phi}{\partial z}=0 & (z=-h), \label{pb:11}
\end{align}
\begin{subequations}\vspace{-2\abovedisplayskip}
\begin{align}
&\displaystyle \frac{\partial \phi}{\partial z}=i\omega \zeta & (z=0), \label{pb:12a}\\
&\displaystyle \zeta=-\frac{i\omega}{g}[1+\psi(X_3^*-1)]\phi & (z=0), \label{pb:12b}\\
&\displaystyle \frac{\partial \phi}{\partial z}=\frac{\omega^2}{g}[1+\psi(X_3^*-1)]\phi& (z=0), \label{pb:12c}
\end{align}
\end{subequations}
where wave amplitude is represented by $\zeta$ as it is a function of the global coordinate.
Here, (\ref{pb:12b}) is the homogenized dynamic condition, and (\ref{pb:12c}) is the homogenized free surface condition.
The result is almost the same as that of \cite{garnaud2009}; only the term of $X_3^*$ is different (except for the definition of phase).

\subsection{Dispersion relation and Wave attenuation in marginal ice zone}
The homogenized boundary value problem (\ref{pb:10}) to (\ref{pb:12c}) is easily solved.
We consider the solution for time-harmonic plane waves denoted by $A={\rm Re}[\zeta \exp(i\omega t)]$ and $\zeta=\zeta(0)\exp(-i\kappa_n x)$ where $\zeta(0)$ is the wave amplitude at the measured up-wave position and $\kappa_n$ is wave number of waves in the marginal ice zone.
Solving the problem, we obtain the dispersion relation as
\begin{align}
\displaystyle \frac{\omega^2}{g}[1+\psi(X_3^*-1)]=\kappa_n\tanh{\kappa_n h}, \label{dis:01}
\end{align}
where the solution of $\kappa_n$ becomes complex because $X_3^*$ is complex.
When the ice floe does not exist on the surface (i.e. the concentration ratio $\psi\to 0$) or the floe is too small (i.e. $X_3^*\to 1$ and  $\arg(X_3^*)\to 0$), (\ref{dis:01}) is deformed as
\begin{align}
\displaystyle \frac{\omega^2}{g}=k_n\tanh{k_n h}, \label{dis:02}
\end{align}
where (\ref{dis:02}) is a dispersion relation of waves on open water and $k_n$ is the wave number of this relation.

\begin{table}
 \caption{Wave numbers for open water $k_n$ and for marginal ice zone $\kappa_n$ at following conditions: circular frequency $\omega=0.5$ rad$/$s, filling ratio $\psi=0.5$, non-dimensional motion amplitude $X_3^*=9.75\times 10^{-1}-3.03i\times 10^{-4}$ (floe radius $a=20$ m and draft $d=1.0$ m), and Water depth is sufficiently deep.}
 \label{table1}
 \par\vspace*{2mm}
 \centering
\begin{tabular}{ccc} 
$n$&$k_n$&$\kappa_n$ \\
0&$\;\;2.55 \;\times 10^{-2}$&$2.52 \times 10^{-2}-3.86i\times 10^{-6}$\\
1&$-7.54i\times 10^{-3}$&$2.01 \times 10^{-7}-7.56i\times 10^{-3}$\\
2&$-2.20i\times 10^{-2}$&$3.39 \times 10^{-7}-2.20i\times 10^{-2}$\\
3&$-3.57i\times 10^{-2}$&$3.09 \times 10^{-7}-3.58i\times 10^{-2}$\\
4&$-4.90i\times 10^{-2}$&$2.61 \times 10^{-7}-4.91i\times 10^{-2}$\\
5&$-6.22i\times 10^{-2}$&$2.21 \times 10^{-7}-6.22i\times 10^{-2}$
\end{tabular}
\end{table}

The example of wave numbers is shown in Table \ref{table1}.
Both wave numbers for open water $k_n$ and the marginal ice zone $\kappa_n$ are calculated at the circular frequency $\omega=0.5$ rad$/$s using (\ref{dis:02}) and (\ref{dis:01}).
Water depth is assumed deep.
The floe size is set as the radius $a=20$ m and draft $d=1.0$ m, and then non-dimensional motion amplitude is given as $X_3^*=9.75\times 10^{-1}-3.03i\times 10^{-4}$.
Furthermore, the concentration ratio $\psi=0.5$ is used.
Table \ref{table1} shows the first six solutions.
Wave number for open water has one real solution $k_0$ and infinite numbers of imaginary solutions $k_1, k_2, \cdots$.
The wave number $k_0$ represents progressive waves, and $k_1, k_2, \cdots$ describe local waves whose amplitudes exponentially decay with distance, respectively.
The wave number for the marginal ice zone has infinite numbers of solutions, however, all solutions are complex.
This indicates that waves propagate with decaying its amplitude; damped waves \cite[e.g.][]{Fox1994} are generated.
Interestingly, the dominant part of $\kappa_n$ (real part for $\kappa_0$ and imaginary part for others) is almost the same as that of $k_n$.
As amplitudes of $\kappa_1, \kappa_2,\cdots$ are small and decaying rapidly, $\kappa_0$ represents the main waves in the marginal ice zone.
It is also indicated that the wavelength is slightly modulated.

Now, we discuss the energy attenuation in the marginal ice zone.
It is known that many field observations indicate energy decays exponentially with distance \cite[e.g.][]{robin1963, wadhams1988}.
Therefore, the energy $E(x)$ may be
\begin{align}
\displaystyle E(x)=E(0)e^{-\alpha x}, \label{dis:03}
\end{align}
where $E(0)$ is initial energy and $\alpha$ is a wave attenuation coefficient.
This can be justified by considering the wave number.
Here, we rewrite the wave number as $\kappa_0=\kappa_{\rm R}-i\kappa_{\rm I}$ where $\kappa_{\rm I}$ is defined positive (for long wave frequencies).
Then, wave elevation becomes
\begin{align}
\displaystyle A(x,t)={\rm Re}[\zeta(0)e^{-i\kappa_0 x}e^{i\omega t}]={\rm Re}[\zeta(0)e^{-\kappa_{\rm I} x}e^{i(\omega t-k_{\rm R}x)}]. \label{dis:04}
\end{align}
Therefore, wave energy is given as
\begin{align}
\displaystyle E(x)\propto |\zeta|^2\propto e^{-2\kappa_{\rm I}x}. \label{dis:05}
\end{align}
This confirms that wave energy exponentially decays with distance, and $\alpha=2\kappa_{\rm I}$.
It should be highlighted that complex wave number and resultant attenuation are known in the dissipation process through two-layer viscous models of an ice plate \cite[]{keller1998, de2002}.
On the other hand, it has not been proven for the scattering process.
In the dissipation process, the imaginary part of the wave number is caused by viscosity; energy dissipates with the distance. 
While this in the scattering process is caused by the floe's motion; energy is conserved, but some wave energies are damped by the floe's motion.
It is also worth noting that incorporating the presence of objects into a dispersion relation is also studied in the field of porous structures \cite[][]{yu1994}.
They often use arrays of the bottom-mounted vertical cylinders \cite[e.g.][]{molin2016,arnaud2017}, and complex wave numbers are given only when the viscous effect is considered.
These studies might be helpful to simulate the wave energy dissipation in the marginal ice zone as the present dissipation process is mainly explained by the boundary layers on the bottom of the ice plate.
We emphasize that boundary layers and resultant vortex at the cylinder's side walls are not always negligible for waves through the array of cylinders \cite[see][]{kagemoto2002, antolloni2020}.
Nevertheless, these are not considered because our purpose is to present the mechanism of wave attenuation by the scattering process.

Using deep water assumption, further discussion is facilitated.
The wave number is then given as
\begin{align}
\displaystyle \kappa_0=\frac{\omega^2}{g}[1+\psi(X_3^*-1)]. \label{de:01}
\end{align}
Therefore, the imaginary part of the wave number is explicitly obtained, and the wave attenuation coefficient becomes
\begin{align}
\displaystyle \alpha=2\kappa_{\rm I}=-2\frac{\omega^2}{g}\psi{\rm Im}[X_3^*]=-2K_0\psi{\rm Im}[X_3^*] \label{de:02}
\end{align}
where $K_0=\omega^2/g$ is the wave number for open water in the deep sea.
This indicates that the wave attenuation coefficient is proportional to wave number $K_0$, concentration ratio $\psi$, and imaginary part of the floe's motion ${\rm Im}[X_3^*]$.

Generally, it is known that the wave attenuation coefficient increases as the wave period becomes shorter \cite[e.g.][]{li2017}.
Our model (\ref{de:02}) also shows this tendency as $\alpha\propto K_0$.
On the other hand, some field observations reported the attenuation coefficient has a peak value at the specific frequency \cite[e.g.][]{wadhams1988, hayes2007, doble2015}; this phenomenon is called the rollover.
Our model may predict such a ``rollover-like phenomenon".
We cannot conclude in the present paper that this is exactly the same phenomenon observed in the fields.
Nevertheless, hereafter we just use ``rollover" for simplification.
This will be discussed later in \S\ref{sec3.1}.
The rollover frequency $\omega_{\rm ro}$ is identified by the condition $\partial \alpha/\partial \omega =0$.
Especially, rollover frequency for deep water is given from (\ref{de:02}) as
\begin{align}
\displaystyle \frac{\partial \alpha}{\partial \omega}=0 \;\to 2{\rm Im}[X_3^*]+\omega \frac{\partial}{\partial \omega}{\rm Im}[X_3^*]=0, \label{de:03}
\end{align}
or we can rewrite it as
\begin{align}
\displaystyle \bigg(2|X^*_3|+\omega\frac{\partial}{\partial \omega}|X_3^*|\bigg)\sin\varepsilon_3+\omega|X_3^*|\frac{\partial \varepsilon_3}{\partial \omega}\cos\varepsilon_3=0, \label{de:04}
\end{align}
where $X_3^*=|X_3^*|\exp(i\varepsilon_3)$ and $\varepsilon_3$ is the phase of the heave motion.
It is obvious that the rollover frequency is independent of the concentration ratio, and the imaginary part of the floe's motion is essential.
For further discussion, we assume the phase is small.
It may be plausible when the floe is thin.
Then, the rollover condition is approximated as
\begin{align}
\displaystyle \frac{\partial \varepsilon_3}{\partial \omega}\approx 0. \label{de:05}
\end{align}
This means that the rollover frequency corresponds to a stationary phase of the motion, and this frequency is only the function of the floe's parameters (i.e. radius and draft).
Both the wave attenuation coefficient and rollover frequency are related to the phase of the floe's motion.
It is known that the reflection and transmission coefficients (for both diffraction and motion-free problems) are described by the phases of the floe \cite[]{newman1975}.
Therefore, it is not surprising that the motion's phase describes the scattering process.

\begin{figure}
     \centering
     \begin{subfigure}{0.497\textwidth}
         \centering
         \includegraphics[width=\textwidth]{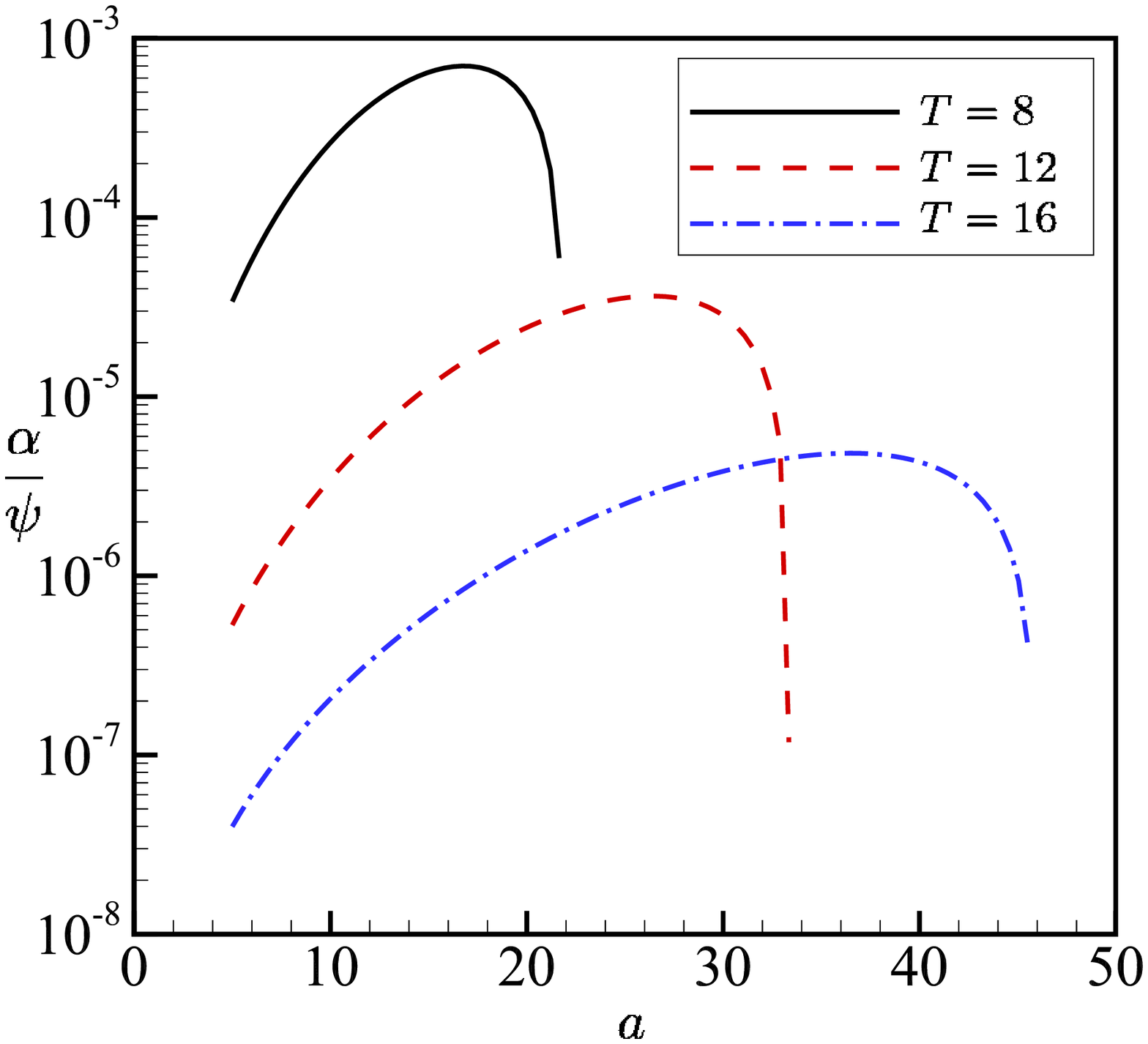}
         \caption{}
        \label{fig4a}
     \end{subfigure}
     \hfill
     \begin{subfigure}{0.497\textwidth}
         \centering
         \includegraphics[width=\textwidth]{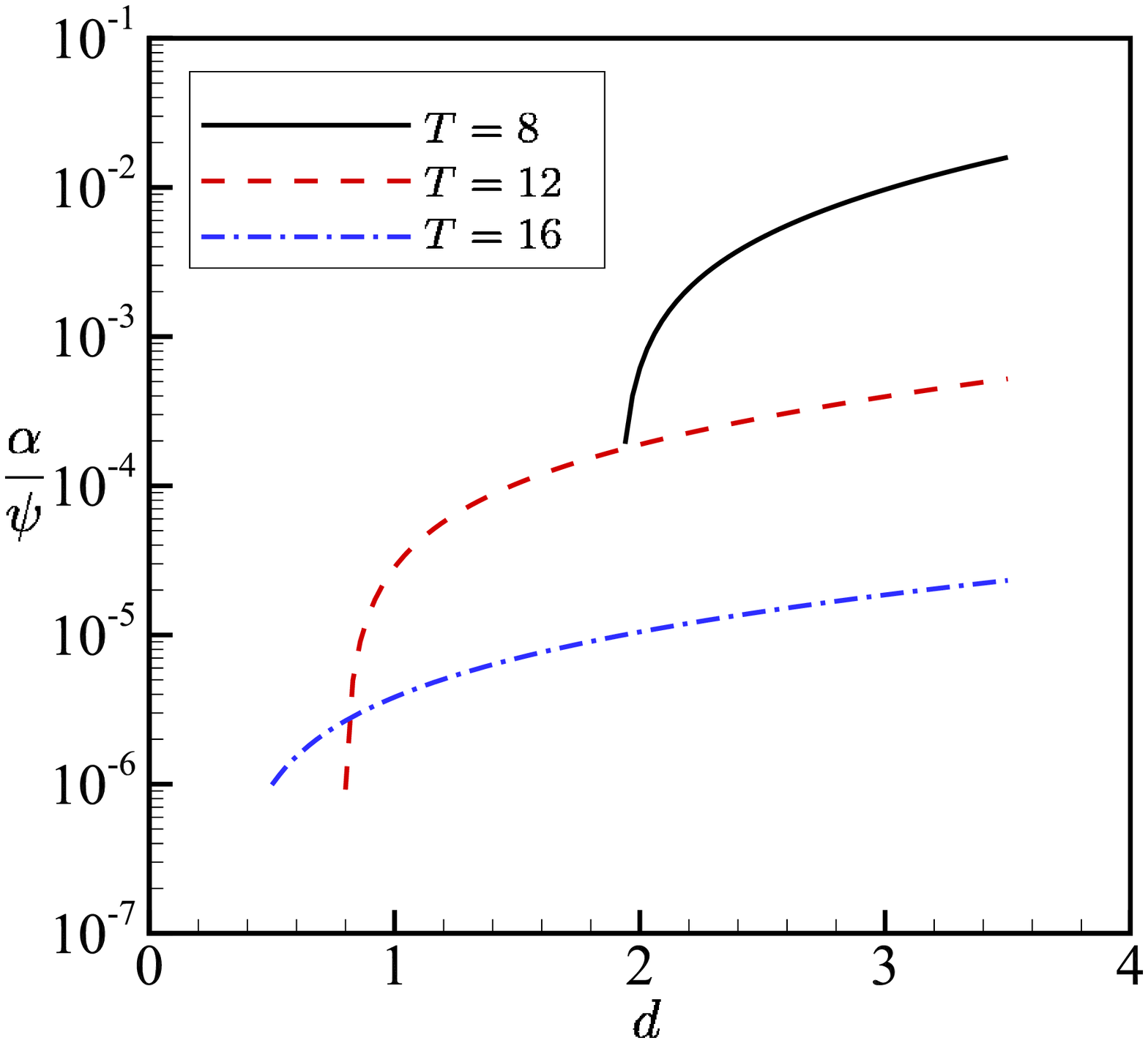}
         \caption{}
        \label{fig4b}
     \end{subfigure}
 \caption{Wave attenuation coefficient against floe's parameters for wave periods $T=8, 12$, and 16 s. The wave attenuation coefficient is normalized by the concentration ratio, i.e. $\alpha/\psi$. Deep water condition is considered. Figures are displayed on semi-log graphs. (a) Coefficients against floe's radius $a$ m. The draft is assumed $d=1.0$ m. (b) Coefficients against floe's draft $d$ m. The radius is assumed $a=30$ m. }
        \label{fig4}
\end{figure}

\section{Simulations and discussion}\label{sec3}

\subsection{Parameter study}\label{sec3.1}

Our model shows a simple but explicit relation between the wave attenuation coefficient and physical parameters as in (\ref{de:02}).
Nevertheless, a numerical simulation is required to obtain the motion of the single ice floe since an analytical solution is not available so far.
Therefore, parameter studies are carried out in this subsection.

As it is obvious from (\ref{de:02}) and (\ref{de:03}), the wave attenuation coefficient is proportional to the concentration ratio $\psi$, and the rollover frequency is independent of the concentration ratio.
Liner relation between the wave attenuation coefficient and concentration ratio is also demonstrated by several papers \cite[e.g.][]{perrie1996, bennetts2010}, and thus this is consistent with them.
Hereafter, we normalize the wave attenuation coefficient by the concentration ratio, i.e. $\alpha/\psi$ is used since the concentration ratio is not important to discuss the influence of floe's parameters.

Normalized wave attenuation coefficients as functions of floe's parameters (radius and draft) are shown in Fig. \ref{fig4}.
Figures are displayed on semi-log graphs, and results of three wave periods $T=8$, 12, and 16 s are plotted.
For these results, deep water condition is assumed.
Figure \ref{fig4a} shows the normalized wave amplitude against the floe's radius $a$.
The draft is kept constant for $d=1.0$ m.
Generally, the wave attenuation coefficient increases as the radius is larger, and shorter wave periods show larger results.
However, the coefficient has a peak at the specific radius, and it falls sharply after the peak.
The existence of the peak implies a rollover-like phenomenon, and this is discussed in the next paragraph.
After the peak, the coefficient drops and becomes negative (although negative values are removed to use the semi-log graph).
Our model assumes the small size of the floe, and the results are not valid for such a big radius.
In addition, the elastic response may not be negligible for these situations.
Figure \ref{fig4b} presents the normalized wave coefficient against the floe's draft $d$.
The radius is constant for $a=30$ m.
Similar to Fig. \ref{fig4a}, the coefficient increases as the draft is larger, and a shorter wave period indicates a larger coefficient.
On the other hand, a drop in coefficient occurs when the draft is small.
Except for drops, qualitative tendencies are well corresponding to other researchers' work \cite[e.g.][]{bennetts2010}.

\begin{figure}
     \centering
     \begin{subfigure}{0.497\textwidth}
         \centering
         \includegraphics[width=\textwidth]{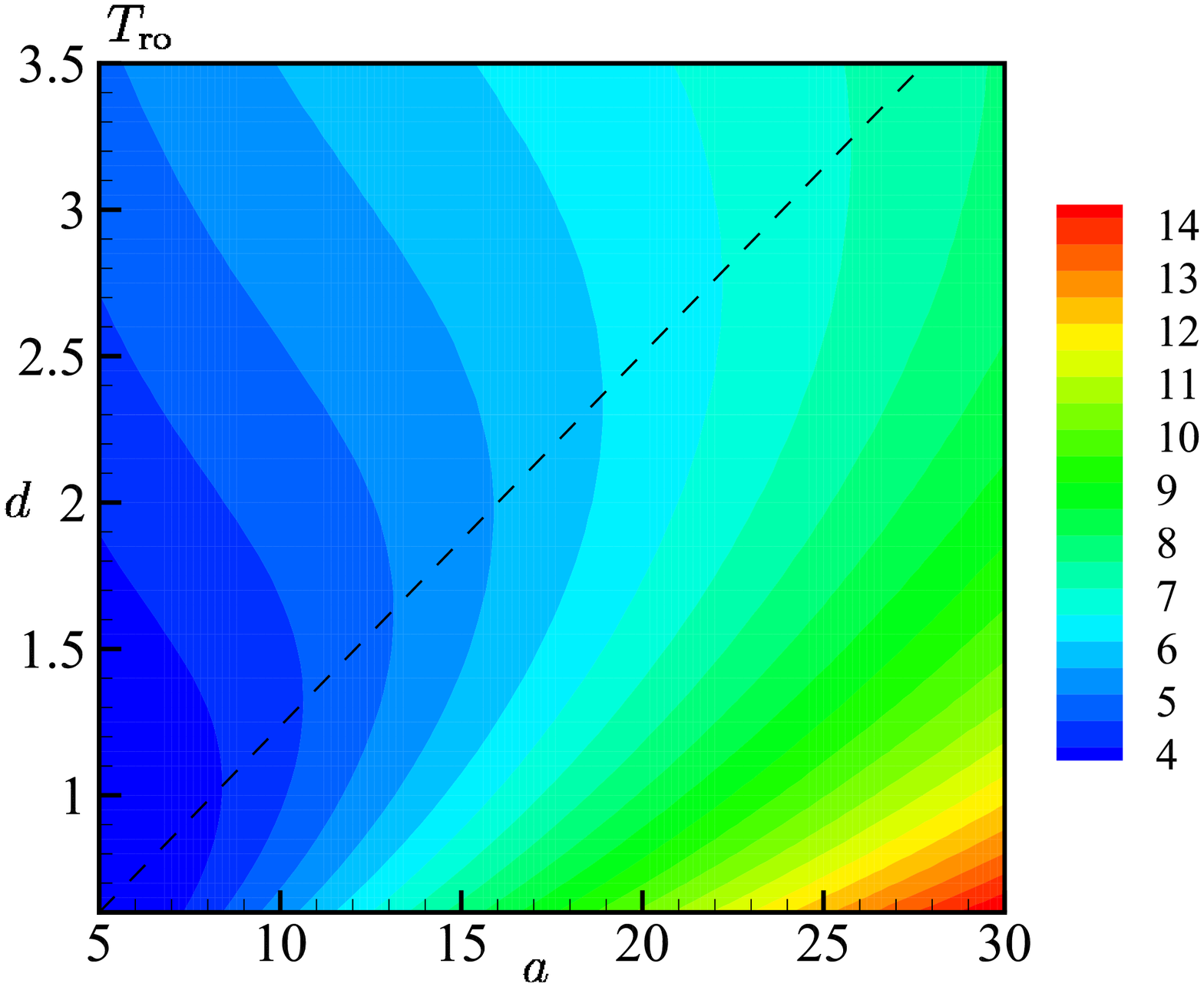}
         \caption{}
        \label{fig2a}
     \end{subfigure}
     \hfill
     \begin{subfigure}{0.497\textwidth}
         \centering
         \includegraphics[width=\textwidth]{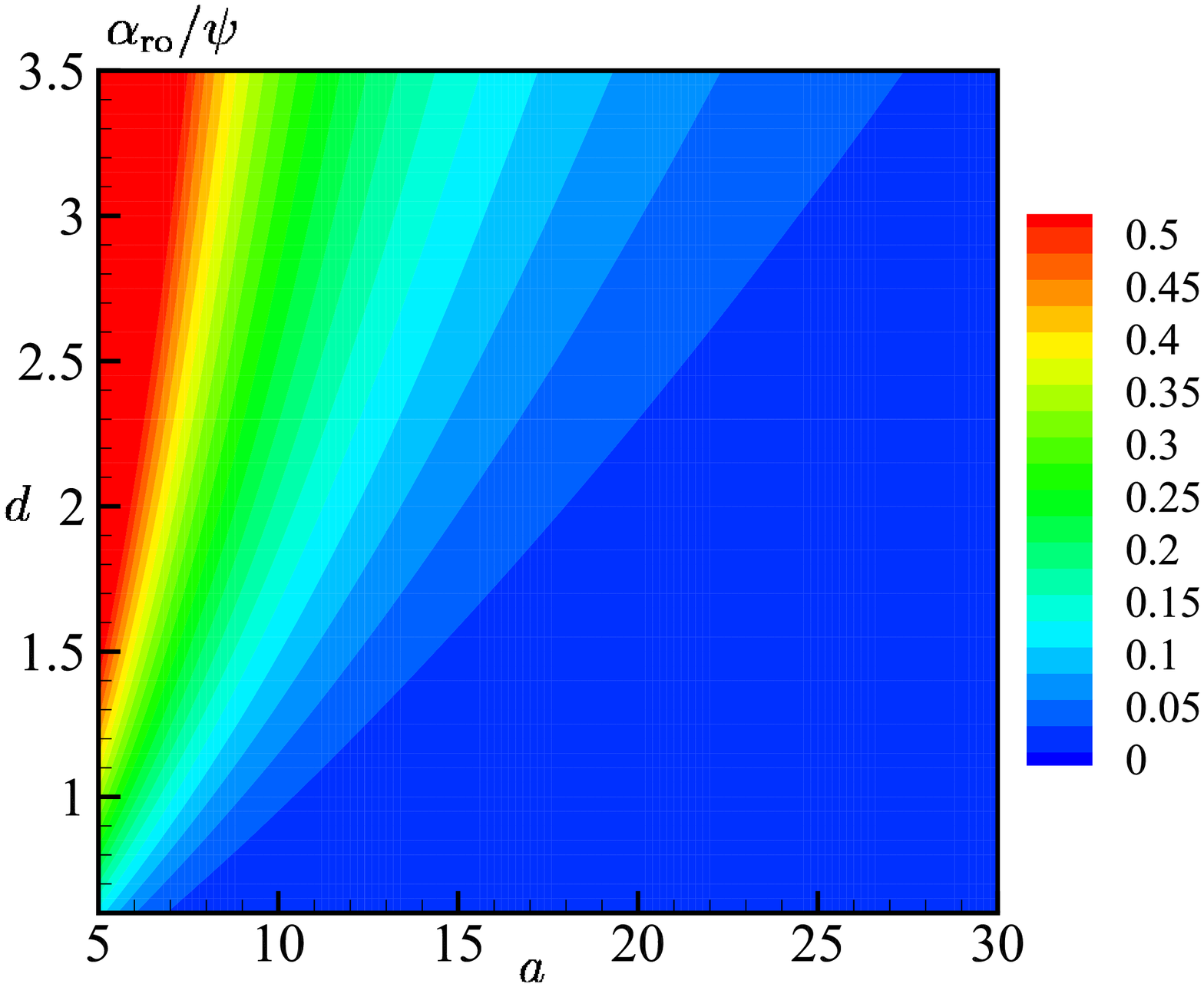}
         \caption{}
        \label{fig2b}
     \end{subfigure}
 \caption{Contour figures related to rollover for deep water waves. (a) Rollover period $T_{\rm ro}$ s vs. floe's parameters, i.e. radius $a$ m and draft $d$ m. Local minima exist for each $a$ at $d\approx 0.126 a$ (dashed line). (b) Normalized wave attenuation coefficient at the rollover period $\alpha_{\rm ro}/\psi$ vs. floe's parameters. }
        \label{fig2}
\end{figure}

Here, the relation between floe's parameters and rollover period $T_{\rm ro}=2\pi/\omega_{\rm ro}$ is shown in Fig. \ref{fig2a} where deep water waves are assumed, and (\ref{de:03}) is used.
Looking at each radius $a$, the rollover period has a local minimum on dashed line $d\approx 0.126 a$.
When $d\le 0.126 a$, strong damping prevents amplitude excitation of the floe's heave motion at the resonance.
We consider such a thin plate, and then the rollover period decreases as the radius becomes smaller or the draft becomes bigger.
The normalized wave attenuation coefficient at the rollover period is shown in Fig. \ref{fig2b}.
This indicates that the wave attenuation coefficient increases as the radius becomes smaller or the draft becomes bigger.

\subsection{Comparisons with field observations and other models}

In order to further demonstrate our model, comparisons with field observation data and other simulation models are carried out.
We choose four fields' data (a) Greenland sea in 1979 \cite[]{wadhams1988}, (b) Bering sea in 1983 \cite[]{wadhams1988}, (c) Bellingshausen sea in 2007 \cite[]{hayes2007}, and (d) Antarctic sea in 2012 \cite[]{kohout2014, li2017} where all fields data show rollover phenomenon at a specific frequency.

Herein, the details of the field observations and the citing simulation models are reviewed.
Firstly, (a) Greenland sea observations were carried out in the King Oscars Fjord area for two days (4th and 10th September 1979) by \cite{wadhams1988}.
On 4th September, the sea was covered by 30\% of multi-year floes, of which typical lengths were 50 to 80 m.
On 10th September, the ice cover was sparser, and larger floes were observed, with a length of 150 m.
The thickness of the ice was estimated to be 3.1 m based on the average of data in King Oscars Fjord for both days.
We also compare with two simulation models.
The first model is based on a two-layer viscous fluid \cite[]{de2002}, tagged as the viscous dissipation model in this paper.
They assumed density of the ice was 922.5 ${\rm kg/m^3}$, and that of fluid was 1025 ${\rm kg/m^3}$.
To assume an infinite depth bottom layer, they used the ice's kinematic viscosity coefficient 1500 to 2000 ${\rm m^2/s}$ and thickness 0.01 to 0.03 m.
In addition, the kinematic viscosity of fluid was tuned as $(1.95\pm 0.73)\times 10^{-3} {\rm m^2/s}$. 
The second one is an elastic plates model \cite[]{kohout2008}, called the elastic scattering model in this paper.
The elastic plate's parameters are Young's modulus of 6.0 GPa, and Poisson's ratio of 0.3.
Both densities of the ice and fluid are the same as in the viscous dissipation model.
We use their result from the 10th September, and they assumed the floe's length of 80 m, the thickness of 3.1 m, and the concentration ratio of 0.17.
Secondly, (b) Bering sea observations were also carried out by \cite{wadhams1988}.
They measured two times with an approximately 12-hour difference on 7th February 1983.
The sea was covered by closely packed brash and floe fragments with concentration ratios of 0.59 to 0.86, and thus they expected the scattering process was more important.
The floe's thicknesses were estimated to be distributed with 20\% of 0.4 m, 40\% of 0.7 m, 0\% of 1.1 m,  and 10\% of 1.7 m.
For the comparison, we quote the results of the energy transfer model \cite[]{perrie1996} as well as the elastic scattering model by \cite{kohout2008} for these experiments.
\cite{perrie1996} used the energy transport equation consisting of terms of wind-wave interaction, nonlinear wave-wave interaction, wave breaking, and wave-ice interaction.
For the wave-ice interaction, they considered the motions of the floe to model the scattering process.
They assumed the following parameters: an ice length of 14.5 m, a thickness of 1.5 m, and a wind speed of 20 ${\rm m/s}$.
On the other hand, \cite{kohout2008} used an ice length of 14.5 m, a thickness of 0.7 m, and a concentration ratio of 0.72.
Note that other parameters are the same as the simulation of (a).
Thirdly, (c) Bellingshausen sea observation was carried out by \cite{hayes2007}.
We use the data from 24th March 2007, and the sea was 100\% covered by ice; 60\% consists of first-year ice floe with less than 20 m lengths and 0.5 to 0.75 m thicknesses, and 40\% consists of brash with 0.5 m thickness.
Therefore, it may not be adequate to simulate this experiment by only the scattering model as the dissipation process is likely to be more important.
Note that \cite{hayes2007} indicated the compromise of data collection for results of wave periods longer than 16 s, and thus we remove these data from the comparison.
A simulation result of the elastic scattering model by \cite{kohout2011} is also used where their model is the same as \cite{kohout2008}.
Although they also proposed a combination model of scattering and drag, we do not use this result because we want to compare the result of the scattering process itself.
They assumed the floe's length of 20 m, the thickness of 0.625 m, and the concentration ratio of 0.6.
This means that the brash ice is ignored.
Finally, (d) Antarctic sea observation was conducted by \cite{kohout2014}, and the data are extracted from \cite{li2017}. 
Observations were carried out between 23th September and 10th October 2012 where the sea was covered mainly by broken first-year ice floes with lengths of 2 to 20 m and thicknesses of 0.5 to 1 m.
Here, we use the median value of a pair of sensors and the mean value of three pairs of sensors in \cite{li2017} for the experimental references.
\cite{li2017} also provided two numerical simulation results.
One is the viscoelastic dissipation model developed by \cite{wang2010}, and the another is the energy transfer model in consideration of the viscoelastic dissipation model.
The simulation parameters were set as Young's ratio of $5.2\times 10^4$ Pa, a viscosity coefficient of 0.2 ${\rm m^2/s}$, an average ice thickness of 0.75 m, and an ice concentration ratio of 0.65.
Note that the median value is extracted for the result of the energy transfer model.

\begin{figure}
     \centering
     \begin{subfigure}{0.497\textwidth}
         \centering
         \includegraphics[width=\textwidth]{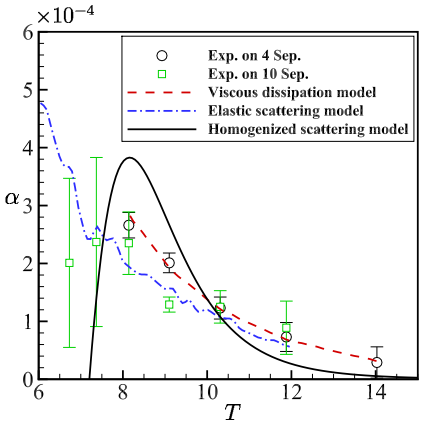}
         \caption{Greenland sea in 1979 \cite[]{wadhams1988}}
        \label{fig3a}
     \end{subfigure}
     \hfill
     \begin{subfigure}{0.497\textwidth}
         \centering
         \includegraphics[width=\textwidth]{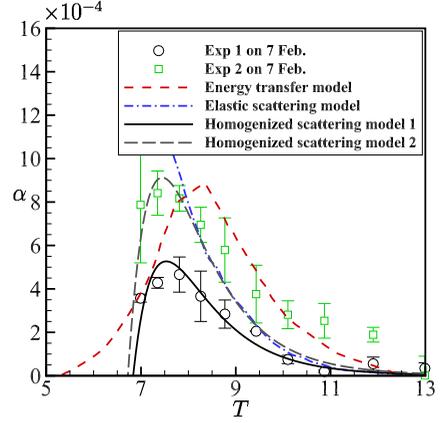}
         \caption{Bering sea in 1983 \\\cite[]{wadhams1988}}
        \label{fig3b}
     \end{subfigure}\\
     \begin{subfigure}{0.497\textwidth}
         \centering
         \includegraphics[width=\textwidth]{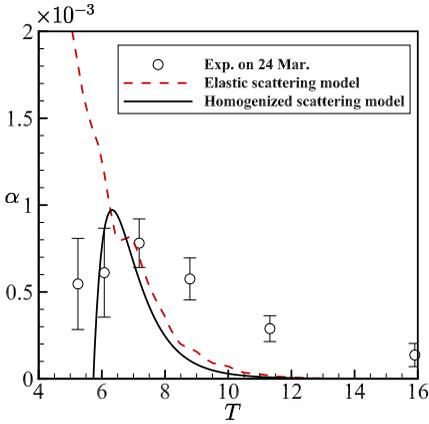}
         \caption{Bellingshausen sea in 2007\\ \cite[]{hayes2007}}
        \label{fig3c}
     \end{subfigure}
     \hfill
     \begin{subfigure}{0.497\textwidth}
         \centering
         \includegraphics[width=\textwidth]{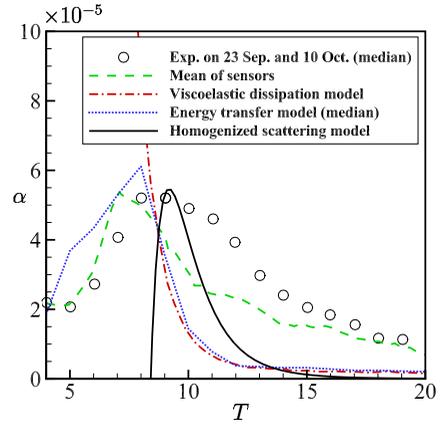}
         \caption{Antarctic sea in 2012\\ \cite[]{kohout2014}}
        \label{fig3d}
     \end{subfigure}
 \caption{Comparisons of wave attenuation coefficient among field observation data and some simulation models. The present results are denoted by the homogenized scattering model. Parameter sets of radius $a$, draft $d$, and concentration ratio $\psi$ used in our simulations are (a) $a=30.0$ m, $d=2.4$ m, and $\psi=0.1$, (b) $a=18.6$ m, $d=1.0$ m, and $\psi=0.72$ for case 1, and $a=20.0$ m, $d=1.2$ m, and $\psi=0.72$ for case 2, (c) $a=13.5$ m, $d=0.75$ m, and $\psi=0.8$, and (d) $a=20.0$ m, $d=0.75$ m, and $\psi=0.65$. 
 Other models are as follows: (a) viscous dissipation model \cite[]{de2002} and elastic scattering model \cite[]{kohout2008},  (b) energy transfer model \cite[]{perrie1996} and elastic scattering model \cite[]{kohout2011}, (c) elastic scattering model \cite[]{kohout2011}, and (d) viscoelastic dissipation model \cite[]{wang2010} (data is obtained from \cite{li2017}) and energy transfer model \cite[]{li2017}. }
        \label{fig3}
\end{figure}

Although some parameters are provided in the references, these have uncertainty because of the complexity of field observations.
In addition, different papers used different values.
Therefore, parameter tuning is inevitable.
It is better to use the experimentally provided data, however, these values may not always yield reasonable results.
In such a case, we choose parameters by the following procedure: (i) the set of the floe's radius and the draft is determined to obtain the almost same rollover period as the experiment, and (ii) the concentration ratio is decided to get the almost same order as the experiment.
As for the simulation of (a) Greenland sea observation, we use the radius $a=30.0$ m (i.e. diameter is 60.0 m), the draft $d=2.4$ m, and the concentration ratio $\psi=0.1$.
For (b) Bering sea observation, two different results were provided, and thus we run two cases of simulations as follows: case 1 uses the radius $a=18.6$ m, the draft $d=1.0$ m, and the concentration ratio $\psi=0.72$, and case 2 uses the radius $a=20.0$ m, the draft $d=1.2$ m, and the concentration ratio $\psi=0.72$.
For (c) Bellingshausen sea observation, the radius $a=13.5$ m, the draft $d=0.75$ m, and the concentration ratio $\psi=0.8$ are used.
Finally, we use the radius $a=20.0$ m, the draft $d=0.75$ m, and the concentration ratio $\psi=0.65$ for (d) Antarctic sea observation.
Note that radii used in (b) and (d) are larger than provided averaged floe lengths because the floe with comparable length has a more important role in wave attenuation rather than the floe with averaged radius.

The wave attenuation coefficients for four field observations are shown in Fig. \ref{fig3}.
The results of the present model are denoted by the homogenized scattering model.
The homogenized scattering model shows the same orders of the wave attenuation coefficients with all field observation data, as well as the rollover-like phenomenon.
When the period is short, the calculated wave attenuation coefficients become negative and thus truncated.
As our model assumes small and rigid ice floes, results are not valid for such a short period.
In fact, even if the attenuation coefficient is positive, the floe's size is comparable to wavelength in some range of periods, including the rollover period.
Especially, the large radius of the floe is used for (a) Greenland sea simulation.
It is reported that the error of the homogenized solution from exact solutions is 1\% for $\ell/\lambda=0.3$, and it still keeps 10\% for $\ell/\lambda=1.0$ \cite[][]{iida2020}, although their study is for a floating thin plate.
Therefore, we expect the present model has at least 10\% error for such a comparable wavelength.

The rollover is not reproduced by elastic scattering, viscous dissipation, and viscoelastic dissipation models. 
On the other hand, the energy transfer models \cite[]{perrie1996, li2017} show such a peak when considering wind and nonlinear energy transfer. 
Despite the field observations, the rollover has not been observed in laboratory experiments so far \cite[]{li2017}. 
Therefore, the interaction of wind and waves is expected a mechanism to explain the rollover \cite[]{li2017}. 
Our model offers a different mechanism; the rollover-like phenomenon occurs because the phase of damping motion of the floe changes in frequency, and the resultant wave reflection changes. 
This is rather a classical viewpoint as already proposed in \cite{wadhams1986}.
\cite{montiel2016} also implied reproduction of such a phenomenon using their scattering model, although they did not conclude this phenomenon is the same as the rollover.
\cite{li2017} pointed out that such reflection-based rollover may be less likely to occur if the variance of floes' size distribution is large. 
Similar to \cite{montiel2016}, we do not conclude this mechanism.
However, it is worth revisiting this viewpoint to understand this phenomenon more deeply.

Interestingly, all models offer similar wave attenuation coefficients despite the physical mechanisms being different (i.e. rigid body-wave interaction, elastic motion, viscous friction, and so on).
These are due in part to parameter tuning.
In the fields, it is hard to imagine that the attenuation process is described by a single mechanism.
It is rather natural to assume this phenomenon is influenced by many factors.
For example, (c) Bellingshausen sea observation could not be explained by only the scattering process because the sea was fully covered by brash, although our result showed a good agreement.
Therefore, the energy transfer model should be developed in corporation with scattering, dissipation, wind-wave interaction, and nonlinear transfer models.

In this paper, we did not discuss the influence of the water depth, and deep water is assumed throughout it.
We expect the water depth effect in the dispersion relation is almost similar to that of open water waves (see (\ref{dis:01}) and (\ref{dis:02})).

\section{Conclusion}
We present a new scattering model to theoretically explain a mechanism of wave attenuation in a marginal ice zone.
To make it simple, we assume the floe is a vertical cylinder, the radius of the floe is small (or at most comparable to wavelength), the draft is also small, the floe is rigid, and the floes are periodically arranged with the same distance.
These assumptions as well as linear potential flow enable us to derive a homogenized free surface condition that is equivalent to the periodic array of ice floes.
The resultant homogenized boundary value problem yields a new dispersion relation, and all wave numbers become complex.
This indicates the exponential decay of wave amplitude and energy with distance.
Under the deep water condition, the wave attenuation coefficient is proportional to the open water's wave number, ice concentration ratio, and imaginary part of the floe's heave motion.
Parameter studies show that the wave attenuation coefficient is bigger as the radius and draft of the floe become bigger.
These tendencies are consistent with the existing research \cite[e.g.][]{bennetts2010} despite our rough assumptions.
Moreover, our model predicts a rollover-like phenomenon; the rollover-like phenomenon occurs at the stationary phase of the motion.
The period of such a phenomenon is bigger as the radius is bigger or the draft is smaller.
Similarly, the peak value is bigger as the radius and draft are bigger.
Comparisons of wave attenuation coefficients against field observations are also demonstrated.
Results are also compared with other simulation models based on various mechanisms.
Our model reproduces similar tendencies with field observations although the scattering is not the only process of wave attenuation.
We believe our theoretical model facilitates the understanding of the scattering process of wave attenuation in the marginal ice zone.

\section*{Acknowledgement }
This work was supported by JSPS KAKENHI Grant Number 21KK0079 and JP19K15218.

\section*{Declaration of interests}
The authors report no conflict of interest.

\bibliographystyle{jfm}
\bibliography{ref}

\begin{thebibliography}{38}
\expandafter\ifx\csname natexlab\endcsname\relax\def\natexlab#1{#1}\fi
\def\au#1{#1} \def\ed#1{#1} \def\yr#1{#1}\def\at#1{#1}\def\jt#1{\textit{#1}}
  \def\bt#1{#1}\def\bvol#1{\textbf{#1}} \def\vol#1{#1} \def\pg#1{#1}
  \def\publ#1{#1}\def\arxiv#1{#1}\def\org#1{#1}\def\st#1{\textit{#1}}

\bibitem[Antolloni {\em et~al.\/}(2020)Antolloni, Jensen, Grue, Riise \&
  Brocchini]{antolloni2020}
{\sc \au{Antolloni, Giulia}, \au{Jensen, Atle}, \au{Grue, John}, \au{Riise,
  Bj{\o}rn~H} \& \au{Brocchini, Maurizio}} \yr{2020}  \at{Wave-induced vortex
  generation around a slender vertical cylinder}.  \jt{Physics of Fluids}
  \bvol{32}~(4),  \pg{042105}.

\bibitem[Arnaud {\em et~al.\/}(2017)Arnaud, Rey, Touboul, Sous, Molin \&
  Gouaud]{arnaud2017}
{\sc \au{Arnaud, Gwendoline}, \au{Rey, Vincent}, \au{Touboul, Julien},
  \au{Sous, Damien}, \au{Molin, Bernard} \& \au{Gouaud, Fabrice}} \yr{2017}
  \at{Wave propagation through dense vertical cylinder arrays: Interference
  process and specific surface effects on damping}.  \jt{Applied Ocean
  Research}  \bvol{65},  \pg{229--237}.

\bibitem[Bennetts \& Squire(2009)]{bennetts2009}
{\sc \au{Bennetts, LG} \& \au{Squire, VA2558590}} \yr{2009}  \at{Wave
  scattering by multiple rows of circular ice floes}.  \jt{Journal of Fluid
  Mechanics}  \bvol{639},  \pg{213--238}.

\bibitem[Bennetts {\em et~al.\/}(2010)Bennetts, Peter, Squire \&
  Meylan]{bennetts2010}
{\sc \au{Bennetts, Luke~G}, \au{Peter, Malte~A}, \au{Squire, VA} \& \au{Meylan,
  Michael~H}} \yr{2010}  \at{A three-dimensional model of wave attenuation in
  the marginal ice zone}.  \jt{Journal of Geophysical Research: Oceans}
  \bvol{115}~(C12).

\bibitem[De~Carolis \& Desiderio(2002)]{de2002}
{\sc \au{De~Carolis, Giacomo} \& \au{Desiderio, Daniela}} \yr{2002}
  \at{Dispersion and attenuation of gravity waves in ice: a two-layer viscous
  fluid model with experimental data validation}.  \jt{Physics Letters A}
  \bvol{305}~(6),  \pg{399--412}.

\bibitem[Doble {\em et~al.\/}(2015)Doble, De~Carolis, Meylan, Bidlot \&
  Wadhams]{doble2015}
{\sc \au{Doble, Martin~J}, \au{De~Carolis, Giacomo}, \au{Meylan, Michael~H},
  \au{Bidlot, Jean-Raymond} \& \au{Wadhams, Peter}} \yr{2015}  \at{Relating
  wave attenuation to pancake ice thickness, using field measurements and model
  results}.  \jt{Geophysical Research Letters}  \bvol{42}~(11),
  \pg{4473--4481}.

\bibitem[Fox \& Squire(1994)]{Fox1994}
{\sc \au{Fox, C.} \& \au{Squire, V.~A.}} \yr{1994}  \at{On the oblique
  reflexion and transmission of ocean waves at shore fast sea ice}.
  \jt{Philosophical Transactions of the Royal Society of London. Series A:
  Physical and Engineering Sciences}  \bvol{347}~(1682),  \pg{185--218}.

\bibitem[Garnaud \& Mei(2009)]{garnaud2009}
{\sc \au{Garnaud, Xavier} \& \au{Mei, Chiang~C}} \yr{2009}  \at{Wave-power
  extraction by a compact array of buoys}.  \jt{Journal of Fluid Mechanics}
  \bvol{635},  \pg{389--413}.

\bibitem[Garrett(1971)]{garrett1971}
{\sc \au{Garrett, CJR}} \yr{1971}  \at{Wave forces on a circular dock}.
  \jt{Journal of Fluid Mechanics}  \bvol{46}~(1),  \pg{129--139}.

\bibitem[Hayes {\em et~al.\/}(2007)Hayes, Jenkins \& McPhail]{hayes2007}
{\sc \au{Hayes, Daniel~R}, \au{Jenkins, Adrian} \& \au{McPhail, Stephen}}
  \yr{2007}  \at{Autonomous underwater vehicle measurements of surface wave
  decay and directional spectra in the marginal sea ice zone}.  \jt{Journal of
  physical oceanography}  \bvol{37}~(1),  \pg{71--83}.

\bibitem[Iida \& Umazume(2020)]{iida2020}
{\sc \au{Iida, Takahito} \& \au{Umazume, Keisuke}} \yr{2020}  \at{Wave response
  of segmented floating plate and validation of its homogenized solution}.
  \jt{Applied Ocean Research}  \bvol{97},  \pg{102083}.

\bibitem[Kagemoto {\em et~al.\/}(2002)Kagemoto, Murai, Saito, Molin {\em
  et~al.\/}]{kagemoto2002}
{\sc \au{Kagemoto, H}, \au{Murai, M}, \au{Saito, M}, \au{Molin, B} \&
  \au{others}} \yr{2002}  \at{Experimental and theoretical analysis of the wave
  decay along a long array of vertical cylinders}.  \jt{Journal of Fluid
  Mechanics}  \bvol{456},  \pg{113--135}.

\bibitem[Keller(1998)]{keller1998}
{\sc \au{Keller, Joseph~B}} \yr{1998}  \at{Gravity waves on ice-covered water}.
   \jt{Journal of Geophysical Research: Oceans}  \bvol{103}~(C4),
  \pg{7663--7669}.

\bibitem[Kohout {\em et~al.\/}(2014)Kohout, Williams, Dean \&
  Meylan]{kohout2014}
{\sc \au{Kohout, AL}, \au{Williams, MJM}, \au{Dean, SM} \& \au{Meylan, MH}}
  \yr{2014}  \at{Storm-induced sea-ice breakup and the implications for ice
  extent}.  \jt{Nature}  \bvol{509}~(7502),  \pg{604--607}.

\bibitem[Kohout \& Meylan(2008)]{kohout2008}
{\sc \au{Kohout, Alison~L} \& \au{Meylan, Michael~H}} \yr{2008}  \at{An elastic
  plate model for wave attenuation and ice floe breaking in the marginal ice
  zone}.  \jt{Journal of Geophysical Research: Oceans}  \bvol{113}~(C9).

\bibitem[Kohout {\em et~al.\/}(2011)Kohout, Meylan \& Plew]{kohout2011}
{\sc \au{Kohout, Alison~L}, \au{Meylan, Michael~H} \& \au{Plew, David~R}}
  \yr{2011}  \at{Wave attenuation in a marginal ice zone due to the bottom
  roughness of ice floes}.  \jt{Annals of Glaciology}  \bvol{52}~(57),
  \pg{118--122}.

\bibitem[Lee \& Lou(1988)]{lee1988}
{\sc \au{Lee, Chih-Kang} \& \au{Lou, Jack~YK}} \yr{1988}  \at{A direct
  boundary-element method for three-d wave diffraction and radiation problems}.
   \jt{Ocean engineering}  \bvol{15}~(5),  \pg{431--455}.

\bibitem[Li {\em et~al.\/}(2017)Li, Kohout, Doble, Wadhams, Guan \&
  Shen]{li2017}
{\sc \au{Li, Jingkai}, \au{Kohout, Alison~L}, \au{Doble, Martin~J},
  \au{Wadhams, Peter}, \au{Guan, Changlong} \& \au{Shen, Hayley~H}} \yr{2017}
  \at{Rollover of apparent wave attenuation in ice covered seas}.  \jt{Journal
  of Geophysical Research: Oceans}  \bvol{122}~(11),  \pg{8557--8566}.

\bibitem[L{\o}ken {\em et~al.\/}(2022)L{\o}ken, Marchenko, Ellevold, Rabault \&
  Jensen]{loken2022}
{\sc \au{L{\o}ken, Trygve~K}, \au{Marchenko, Aleksey}, \au{Ellevold, Thea~J},
  \au{Rabault, Jean} \& \au{Jensen, Atle}} \yr{2022}  \at{Experiments on
  turbulence from colliding ice floes}.  \jt{Physics of Fluids}  \bvol{34}~(6),
   \pg{065133}.

\bibitem[Mei(2012)]{mei2012}
{\sc \au{Mei, Chiang~C}} \yr{2012}  \at{Hydrodynamic principles of wave power
  extraction}.  \jt{Philosophical Transactions of the Royal Society A:
  Mathematical, Physical and Engineering Sciences}  \bvol{370}~(1959),
  \pg{208--234}.

\bibitem[Miles \& Gilbert(1968)]{miles1968}
{\sc \au{Miles, John} \& \au{Gilbert, Freeman}} \yr{1968}  \at{Scattering of
  gravity waves by a circular dock}.  \jt{Journal of Fluid Mechanics}
  \bvol{34}~(4),  \pg{783--793}.

\bibitem[Molin {\em et~al.\/}(2016)Molin, Remy, Arnaud, Rey, Touboul \&
  Sous]{molin2016}
{\sc \au{Molin, Bernard}, \au{Remy, Fabien}, \au{Arnaud, Gwendoline}, \au{Rey,
  Vincent}, \au{Touboul, Julien} \& \au{Sous, Damien}} \yr{2016}  \at{On the
  dispersion equation for linear waves traveling through or over dense arrays
  of vertical cylinders}.  \jt{Applied Ocean Research}  \bvol{61},
  \pg{148--155}.

\bibitem[Montiel {\em et~al.\/}(2016)Montiel, Squire \& Bennetts]{montiel2016}
{\sc \au{Montiel, Fabien}, \au{Squire, VA} \& \au{Bennetts, LG}} \yr{2016}
  \at{Attenuation and directional spreading of ocean wave spectra in the
  marginal ice zone}.  \jt{Journal of Fluid Mechanics}  \bvol{790},
  \pg{492--522}.

\bibitem[Newman(1975)]{newman1975}
{\sc \au{Newman, John~Nicholas}} \yr{1975}  \at{Interaction of waves with
  two-dimensional obstacles: a relation between the radiation and scattering
  problems}.  \jt{Journal of Fluid Mechanics}  \bvol{71}~(2),  \pg{273--282}.

\bibitem[Newman(2018)]{newman2018}
{\sc \au{Newman, John~Nicholas}} \yr{2018} {\em Marine hydrodynamics\/}.
  \publ{The MIT Press}.

\bibitem[Perrie \& Hu(1996)]{perrie1996}
{\sc \au{Perrie, W} \& \au{Hu, Y}} \yr{1996}  \at{Air--ice--ocean momentum
  exchange. part 1: Energy transfer between waves and ice floes}.  \jt{Journal
  of physical oceanography}  \bvol{26}~(9),  \pg{1705--1720}.

\bibitem[Robin(1963)]{robin1963}
{\sc \au{Robin, G de~Q}} \yr{1963}  \at{Wave propagation through fields of pack
  ice}.  \jt{Philosophical Transactions of the Royal Society of London. Series
  A, Mathematical and Physical Sciences}  \bvol{255}~(1057),  \pg{313--339}.

\bibitem[Squire(2020)]{squire2020}
{\sc \au{Squire, Vernon~A}} \yr{2020}  \at{Ocean wave interactions with sea
  ice: A reappraisal}.  \jt{Annual Review of Fluid Mechanics}  \bvol{52},
  \pg{37--60}.

\bibitem[Squire {\em et~al.\/}(1995)Squire, Dugan, Wadhams, Rottier \&
  Liu]{squire1995}
{\sc \au{Squire, Vernon~A}, \au{Dugan, John~P}, \au{Wadhams, Peter},
  \au{Rottier, Philip~J} \& \au{Liu, Antony~K}} \yr{1995}  \at{Of ocean waves
  and sea ice}.  \jt{annual review of fluid mechanics}  \bvol{27}~(1),
  \pg{115--168}.

\bibitem[Stroeve {\em et~al.\/}(2007)Stroeve, Holland, Meier, Scambos \&
  Serreze]{stroeve2007}
{\sc \au{Stroeve, Julienne}, \au{Holland, Marika~M}, \au{Meier, Walt},
  \au{Scambos, Ted} \& \au{Serreze, Mark}} \yr{2007}  \at{Arctic sea ice
  decline: Faster than forecast}.  \jt{Geophysical research letters}
  \bvol{34}~(9).

\bibitem[Sutherland {\em et~al.\/}(2019)Sutherland, Rabault, Christensen \&
  Jensen]{sutherland2019}
{\sc \au{Sutherland, Graig}, \au{Rabault, Jean}, \au{Christensen, Kai~H} \&
  \au{Jensen, Atle}} \yr{2019}  \at{A two layer model for wave dissipation in
  sea ice}.  \jt{Applied Ocean Research}  \bvol{88},  \pg{111--118}.

\bibitem[Suzuki {\em et~al.\/}(1996)Suzuki, Yoshida \& Iijima]{suzuki1996}
{\sc \au{Suzuki, Hideyuki}, \au{Yoshida, Koichiro} \& \au{Iijima, Kazuhiro}}
  \yr{1996}  \at{A consideration of the structural design of a large-scale
  floating structure}.  \jt{Journal of marine science and technology}
  \bvol{1}~(5),  \pg{255--267}.

\bibitem[Thomson {\em et~al.\/}(2018)Thomson, Ackley, Girard-Ardhuin, Ardhuin,
  Babanin, Boutin, Brozena, Cheng, Collins, Doble {\em et~al.\/}]{thomson2018}
{\sc \au{Thomson, Jim}, \au{Ackley, Stephen}, \au{Girard-Ardhuin, Fanny},
  \au{Ardhuin, Fabrice}, \au{Babanin, Alex}, \au{Boutin, Guillaume},
  \au{Brozena, John}, \au{Cheng, Sukun}, \au{Collins, Clarence}, \au{Doble,
  Martin} \& \au{others}} \yr{2018}  \at{Overview of the arctic sea state and
  boundary layer physics program}.  \jt{Journal of Geophysical Research:
  Oceans}  \bvol{123}~(12),  \pg{8674--8687}.

\bibitem[Wadhams(1986)]{wadhams1986}
{\sc \au{Wadhams, Peter}} \yr{1986}  \at{The seasonal ice zone}.  \bt{In {\em
  The geophysics of sea ice\/}},  \pg{pp. 825--991}.  \publ{Springer}.

\bibitem[Wadhams {\em et~al.\/}(1988)Wadhams, Squire, Goodman, Cowan \&
  Moore]{wadhams1988}
{\sc \au{Wadhams, Peter}, \au{Squire, Vernon~A}, \au{Goodman, Dougal~J},
  \au{Cowan, Andrew~M} \& \au{Moore, Stuart~C}} \yr{1988}  \at{The attenuation
  rates of ocean waves in the marginal ice zone}.  \jt{Journal of Geophysical
  Research: Oceans}  \bvol{93}~(C6),  \pg{6799--6818}.

\bibitem[Wang \& Shen(2010)]{wang2010}
{\sc \au{Wang, Ruixue} \& \au{Shen, Hayley~H}} \yr{2010}  \at{Gravity waves
  propagating into an ice-covered ocean: A viscoelastic model}.  \jt{Journal of
  Geophysical Research: Oceans}  \bvol{115}~(C6).

\bibitem[WW3DG(2019)]{wavewatch2019}
{\sc \au{WW3DG}} \yr{2019}  \at{User manual and system documentation of
  wavewatch iii{\textregistered} version 6.07}.  \jt{Technical note}
  \bvol{333},  \pg{1--465}.

\bibitem[Yu \& Chwang(1994)]{yu1994}
{\sc \au{Yu, Xiping} \& \au{Chwang, Allen~T}} \yr{1994}  \at{Wave motion
  through porous structures}.  \jt{Journal of engineering mechanics}
  \bvol{120}~(5),  \pg{989--1008}.

\end{thebibliography}

\end{document}